\documentclass[a4paper,canadian]{article}
\usepackage[T1]{fontenc}
\usepackage[latin9]{inputenc}
\usepackage{verbatim}
\usepackage{url}
\usepackage{amsmath}
\usepackage{graphicx}

\makeatletter

\providecommand{\tabularnewline}{\\}

\usepackage{eusipco2009}
\usepackage{epsfig}
\usepackage{algorithmic}

\title{TATA}
\name{Author(s) Name(s)\thanks{Thanks to XYZ agency for funding.}}
\address{Author Affiliation(s)}
\makeatother

\usepackage{babel}
\begin{document}

\name{Jean-Marc Valin$^{\dag\ddag}$\thanks{We would like to thank all the listeners who participated in the subjective evaluation.}, Timothy B. Terriberry$^\ddag$, Gregory Maxwell$^\ddag$}

\address{$^\dag$Octasic Semiconductor \\
4101 Molson Street, suite 300 \\
Montreal (Quebec) Canada \\
jean-marc.valin@octasic.com \\
$^\ddag$Xiph.Org Foundation \\
tterribe@xiph.org\\
greg@xiph.org}

\title{A Full-Bandwidth Audio Codec with Low Complexity and Very Low Delay}
\maketitle
\begin{abstract}
We propose an audio codec that addresses the low-delay requirements
of some applications such as network music performance. The codec
is based on the modified discrete cosine transform (MDCT) with very
short frames and uses gain-shape quantization to preserve the spectral
envelope. The short frame sizes required for low delay typically hinder
the performance of transform codecs. However, at 96~kbit/s and with
only 4~ms algorithmic delay, the proposed codec out-performs the
ULD codec operating at the same rate. The total complexity of the
codec is small, at only 17~WMOPS for real-time operation at 48~kHz. 
\end{abstract}

\section{Introduction}

\label{sec:intro}

Recent research has focused on increasing the audio quality of speech
codecs to ``full bandwidth'' rates of 44.1 or 48~kHz to make them
suitable to more general purpose applications~\cite{Lutzky2005,Siren14}.
However, while such codecs have moderate algorithmic delays, some
applications require very low delay. One example is networked music
performance, where two or more musicians playing remotely require
less than 25~ms of total delay to be able to properly synchronize
with each other~\cite{Carot2006}. Another example is a wireless
audio device, such as a digital microphone, where delay causes desynchronization
with the visible speaker. Teleconferencing systems where only limited
acoustic echo control is possible also benefit from very low delay,
as it makes acoustic echo less perceptible.

We propose a codec that provides high audio quality while maintaining
very low delay. Its characteristics are as follows:
\begin{itemize}
\item sampling rate of 48 kHz;
\item frame size of 256 samples (5.3 ms) with 128 samples look-ahead (2.7
ms);
\item achieves very good audio quality at 64 kbit/s (mono);
\item a total complexity of 17 WMOPS;
\item optional support for other sampling rates and frame sizes, such as
128-sample frames with 64 samples look-ahead%
.
\end{itemize}
We introduce the basic principles of the codec in Section~\ref{sec:overview}
and go into the details of the quantization in Section~\ref{sec:Quantisation}.
We then discuss how the proposed approach compares to other low-delay
codecs in Section~\ref{sec:related-work}, followed by direct audio
quality comparisons in Section~\ref{sec:results} and the conclusion
in Section~\ref{sec:conclusion}.

\section{Overview of the codec}

\label{sec:overview}

\begin{figure}
\centering\includegraphics[clip,width=1\columnwidth]{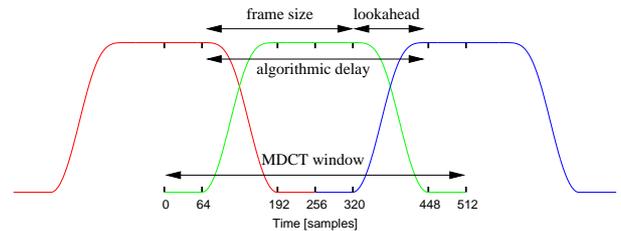}

\caption{Power-complementary windows with reduced overlap.\label{fig:Power-complementary-windows}}
\end{figure}

The proposed codec is based on the modified discrete cosine transform
(MDCT). To minimize the algorithmic delay, we use a short frame size,
combined with a reduced-overlap window. This results in an algorithmic
delay of 384 samples for the 256-sample frame size configuration shown
in Fig. \ref{fig:Power-complementary-windows}.

\begin{figure}
\centering\includegraphics[width=1\columnwidth]{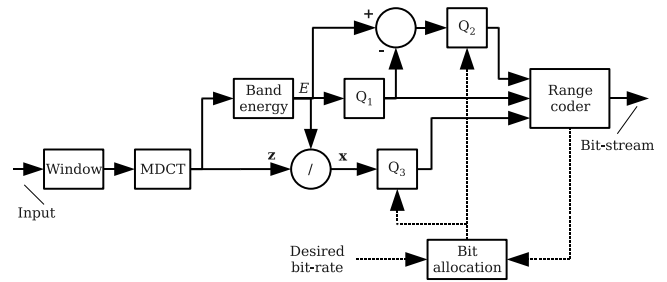}

\caption{Basic structure of the encoder.\label{fig:encoder-structure}}
\end{figure}

The structure of the encoder is shown in Fig. \ref{fig:encoder-structure}
and its basic principles can be summarized as follows:%

\begin{itemize}
\item the MDCT output is split in bands approximating the critical bands;
\item the energy (gain) in each band is quantized and transmitted separately;
\item the details (shape) in each band are quantized algebraically using
a spherical codebook;
\item the bit allocation is inferred from information shared between the
encoder and the decoder.
\end{itemize}

The most important aspect of these is the explicit coding of a per-band
energy constraint combined with an independent shape quantizer which
never violates that constraint. This prevents artifacts caused by
energy collapse or overshoot and preserves the spectral envelope's
evolution in time. The bands are defined to match the ear's critical
bands as closely as possible, with the restriction that bands must
be at least 3 MDCT bins wide. This lower limit results in 19 bands
for the codec when 256-sample frames are used.

\section{Quantization}

\label{sec:Quantisation}

We use a type of arithmetic coder called a range coder~\cite{Mart79}
for all symbols. We use it not only for entropy coding, but also to
approximate the infinite precision arithmetic required to optimally
encode integers whose range is not a power of two.

\subsection{Energy quantization ($Q_{1}$, $Q_{2}$)}

The energy of the final decoded signal is algebraically constrained
to match the explicitly coded energy exactly. Therefore it is important
to quantize the energy with sufficient resolution because later stages
cannot compensate for quantization error in the energy. It is perceptually
important to preserve the band energy, regardless of the resolution
used for encoding the band shape.

We use a coarse-fine strategy for encoding the energy in the log domain
(dB). The coarse quantization of the energy ($Q_{1}$) uses a fixed
resolution of 6~dB. This is also the only place we use prediction
and entropy coding. The prediction is applied both in time (using
the previous frame) and in frequency (using the previous band). The
2-D \emph{z}-transform of the prediction filter is
\begin{equation}
A\left(z_{\ell},z_{b}\right)=\left(1-\alpha z_{\ell}^{-1}\right)\cdot\frac{1-z_{b}^{-1}}{1-\beta z_{b}^{-1}}\ ,
\end{equation}
where $b$ is the band index and $\ell$ is the frame index. Unlike
methods which require predictor reset~\cite{Wabnik2005Packet}, the
proposed system with $\alpha<1$ is guaranteed to re-synchronize after
a transmission error. %
{} We have obtained good results with $\alpha=0.8$ and $\beta=0.7$.
To prevent error accumulation, the prediction is applied on the quantized
log-energy. The prediction step reduces the entropy of the coarsely-quantized
energy from 61 to 30~bits. Of this 31-bit reduction, 12 are due to
inter-frame prediction. We approximate the ideal probability distribution
of the prediction error using a Laplace distribution, which results
in an average of 33~bits per frame to encode the energy of all 19
bands at a 6~dB resolution. Because of the short frames, this represents
up to a 16\% %
bitrate savings on the configurations tested in Section~\ref{sec:results}.

The fine energy quantizer ($Q_{2}$) is applied to the $Q_{1}$ quantization
error and has a variable resolution that depends on the bit allocation.
We do not use entropy coding for $Q_{2}$ since the quantization error
of $Q_{1}$ is mostly white and uniformly distributed.

\subsection{Shape quantization ($Q_{3}$)}

We normalize each band by the unquantized energy, so its shape always
has unit norm. We thus need to quantize a vector on the surface of
an $N$-dimensional hyper-sphere.

\subsubsection{Pyramid vector quantization of the shape}

\label{sub:pvq}

Because there is no known algebraic formulation for the optimal tessellation
of a hyper-sphere of arbitrary dimension $N$, we use a codebook constructed
as the sum of $K$ signed unit pulses normalized to the surface of
the hyper-sphere. A codevector $\tilde{\mathbf{x}}$ can be expressed
as: 
\begin{align}
\mathbf{y} & =\sum_{k=1}^{K}s^{(k)}\varepsilon_{n^{(k)}}\ ,\\
\tilde{\mathbf{x}} & =\frac{\mathbf{y}}{\sqrt{\mathbf{y}^{T}\mathbf{y}}}\ ,
\end{align}
where $n^{(k)}$ and $s^{(k)}$ are the position and sign of the $k^{th}$
pulse, respectively, and $\varepsilon_{n^{(k)}}$ is the $n^{(k)}$th
elementary basis vector. The signs $s_{k}$ are constrained such that
$n^{(j)}=n^{(k)}$ implies $s^{(j)}=s^{(k)}$, and hence $\mathbf{y}$
satisfies $\left\Vert \mathbf{y}\right\Vert _{L1}=K$. This codebook
has the same structure as the pyramid vector quantizer~\cite{Fisher1986}
and is similar to that used in many ACELP-based~\cite{Laflamme1990}
speech codecs.

The search for the best positions and signs is based on minimizing
the cost function $J=-\mathbf{x}\tilde{\mathbf{x}}=-\mathbf{x}\mathbf{y}/\sqrt{\mathbf{y}^{T}\mathbf{y}}$
using a greedy search, one pulse at a time. For iteration $k$, the
cost $J_{n}^{\left(k\right)}$ of placing a pulse at position $n$
can be computed as: 
\begin{align}
s^{(k)}= & \mathrm{sign}\left(x_{n^{(k)}}\right)\ ,\label{eq:cost-sign}\\
R_{xy}^{\left(k\right)}= & R_{xy}^{(k-1)}+s^{(k)}x_{n^{(k)}}\ ,\\
R_{yy}^{\left(k\right)}= & R_{yy}^{(k-1)}+2s^{(k)}y_{n^{(k)}}^{(k-1)}+\left(s^{(k)}\right)^{2}\ ,\\
J_{n}^{\left(k\right)}= & -\left(R_{xy}^{\left(k\right)}\right)^{2}/R_{yy}^{(k)}\ .\label{eq:cost-j-div}
\end{align}
To compare two costs, the divisions in \eqref{eq:cost-j-div} are
transformed into two multiplications. The algorithm can be sped up
by starting from a projection of $\mathbf{x}$ onto the pyramid, as
suggested in~\cite{Fisher1986}. We start from
\begin{equation}
y_{n}=\left\lfloor K\frac{x_{n}}{\left\Vert \mathbf{x}\right\Vert _{L1}}\right\rfloor \ ,
\end{equation}
where $\left\lfloor \cdot\right\rfloor $ denotes rounding towards
zero. From there, we add any remaining pulses one at a time using
the search procedure described by \eqref{eq:cost-sign}-\eqref{eq:cost-j-div}.
The worst-case complexity is thus $O\left(N\cdot\min\left(N,K\right)\right)$.

\subsubsection{Encoding of the pulses' signs and positions}

For $K$ pulses in $N$ samples, the number of codebook entries is
\begin{multline}
V\left(N,K\right)=V\left(N-1,K\right)+\\
V\left(N,K-1\right)+V\left(N-1,K-1\right)\ ,\label{eq:V_NK}
\end{multline}
with $V\left(N,0\right)=1$ and $V\left(0,K\right)=0,\ K>0$. We use
an enumeration algorithm to convert between codebook entries and integers
between 0 and $V\left(N,K\right)-1$~\cite{Fisher1986}. The index
is encoded with the range coder using equiprobable symbols. The factorial
pulse coding (FPC) method~\cite{ACMP00} uses the same codebook with
a different enumeration. However, it requires multiplications and
divisions, whereas ours can be implemented using only addition. To
keep computational complexity low, the band is recursively partitioned
in half when the size of the codebook exceeds 32 bits. The number
of pulses in the first half is explicitly encoded, and then each half
of the vector is coded independently. %

\subsubsection{Avoiding sparseness}

When a band is allocated few bits, the codebook described in section~\ref{sub:pvq}
produces a sparse spectrum, containing only a few non-zero values.
This tends to produce ``birdie'' artifacts, common to many transform
codecs. To mitigate the problem, we add some small values to the spectrum.
We could use a noise generator, but choose to use a scaled copy of
the lower frequency MDCT bins. Doing so mostly preserves the temporal
aspect of the signal~\cite{Makhoul1979}. The gain applied is computed
as:
\begin{equation}
g=\frac{N}{N+\delta K}\ ,\label{eq:fold-gain}
\end{equation}
where $\delta=6$ was experimentally found to be a good compromise
between excessive noise and a sparse spectrum. The gain in~\eqref{eq:fold-gain}
increases as fewer pulses are used. For cases where no pulse is allocated,
we have $g=1$, which preserves the energy in the band without using
any additional bits. In all cases, the total energy is normalized
to be equal to the energy value encoded. This constraint slightly
changes the objective function used to place pulses, but for simplicity
we only take this into consideration when placing the last pulse.

\subsubsection{Avoiding pre-echo}

Pre-echo is a common artifact in transform codecs, introduced because
quantization error is spread over an entire window, including samples
before a transient event. It is seldom a problem in the proposed codec
because of the short frames, but occurs in some extreme cases. To
avoid pre-echo, we detect transients and use two smaller MDCTs for
those frames. The output of the two MDCTs is interleaved, and the
rest of the codec is not affected, operating as if only one MDCT was
used. No additional lookahead is needed to determine which window
to use because the long windows have the same window overlap shape
and length as the short windows.

\subsection{Bit allocation}

The shorter the frame size used in a codec, the higher the overhead
of transmitting metadata. In low-delay codecs, the overhead of explicitly
transmitting the bit allocation can become very large. For this reason,
we choose not to transmit the bit allocation explicitly, but rather
derive it using information available to both the encoder and the
decoder. 

We assume that both the encoder and the decoder know how many 8-bit
bytes are used to encode a frame. This number is either agreed on
when establishing the communication or obtained during the communication,
e.g. the decoder knows the size of any UDP datagram it receives. After
determining the number of bits used by the coarse quantization of
the energy ($Q_{1}$), both the encoder and decoder make an initial
bit allocation for the fine energy ($Q_{2}$) and shape ($Q_{3}$)
using only static (ROM) data. Because we cannot always choose a pulse
count $K$ that yields exactly the number of bits desired for a band,
we use the closest possible value and propagate the difference to
the remaining bands. 

For a given band, the bit allocation is nearly constant across frames
that use the same number of bits for $Q_{1}$, yielding a pre-defined
signal-to-mask ratio (SMR) for each band. Because the bands have a
width of one Bark, this is equivalent to modeling the masking occurring
within each critical band, while ignoring inter-band masking and tone-vs-noise
characteristics. This is not an optimal bit allocation, but it provides
good results without requiring the transmission of any allocation
information. The average bit allocation between the three quantizers
is given in Table~\ref{tab:Average-bit-allocation}.

\begin{table}
\caption{Average bit allocation at 64.5~kbit/s (344 bits per frame). The mode
flags are used for pre-echo avoidance and to signal the low-complexity
mode described here.\label{tab:Average-bit-allocation}}

\centering%
\begin{tabular}{cc}
\hline 
Parameter & Average bits\tabularnewline
\hline 
Coarse energy ($Q_{1}$) & 32.8\tabularnewline
Fine energy ($Q_{2}$) & 43.2\tabularnewline
Shape ($Q_{3}$) & 264.4\tabularnewline
Mode flags & 2\tabularnewline
Unallocated & 1.6\tabularnewline
\hline 
\end{tabular}
\end{table}

\section{Related Work}

\label{sec:related-work}

The proposed codec shares some similarities with the G.722.1C~\cite{Siren14}
audio codec in that both transmit the energy of MDCT bands explicitly.
There are, however, significant algorithmic differences between the
two codecs. First, G.722.1C uses scalar quantization to encode the
normalized spectrum in each band, so it must encode $N$ degrees of
freedom instead of the $N-1$ required by a spherical codebook. For
a Gaussian source, pyramid vector quantization provides a 2.39~dB
asymptotic improvement over the optimal scalar quantizer, according
to~\cite{Fisher1986}. We replaced the VQ codebook in our codec with
per-bin entropy coding, and measured a 10~kbit/s degredation from
our 64~kbit/s configuration. The use of scalar quantization also
means that the energy is not guaranteed to be preserved in the decoded
signal.

\begin{table*}
\caption{Characteristics of the codecs as used in testing\label{tab:Characteristics-of-codecs}}
\centering%
\begin{tabular}{lccrrr}
\hline 
Codec & Sample rate & Bitrate  & Frame size & Look-ahead & \textbf{Total delay}\tabularnewline
 & kHz & kbit/s & sample (ms) & sample (ms) & sample (\textbf{ms})\tabularnewline
\hline 
Proposed (64) & 48 & 64 & 256~(5.3) & 128~(2.7) & 384~~~~(\textbf{8})\tabularnewline
Proposed (96) & 48 & 96 & 128~(2.7) & 64~(1.3) & 192~~~~(\textbf{4})\tabularnewline
ULD & 48 & 96 & 128~(2.7) & 128~(2.7) & 256~(\textbf{5.3})\tabularnewline
G.722.1C & 32 & 48 & 640~~(20) & 640~~(20) & 1280~~(\textbf{40})\tabularnewline
\hline 
\end{tabular}
\end{table*}

A second difference is that in G.722.1C, the bit allocation information
is explicitly transmitted in the bitstream. Given that G.722.1C has
20~ms frames, this is a reasonable strategy. However, with the very
short frames (5.8~ms or less) used in the proposed codec, explicitly
encoding the bit allocation in each frame would result in too much
overhead. A third difference is that the bands in G.722.1C have a
fixed width of 500~Hz. While this helps reduce the complexity of
the codec, which is around 11~WMOPS at 48~kbit/s, it has a cost
in quality compared to using Bark-spaced bands. Besides the differences
in the core algorithm, G.722.1C has a lower complexity and a significantly
larger (5 to 10 times) delay than the proposed codec, so the potential
set of applications for the two codecs only partially overlap.

The Fraunhofer Ultra Low Delay (ULD) codec~\cite{Schuller2002} is
one of the only full-bandwidth codecs with an algorithmic delay comparable
to the proposed codec. Its structure, however, is completely different
from that of the proposed codec. ULD is based on time-domain linear
prediction instead of the MDCT. It uses a pre-filter/post-filter combination,
whose parameters are transmitted in the bitstream, to shape the quantization
noise. ULD frames are 128 samples with 128 samples of look-ahead,
for a total algorithmic delay of 256 samples at 48~kHz (5.3~ms).
One disadvantage of the linear-prediction approach is the difficulty
of resynchronizing the decoder after a packet is lost~\cite{Wabnik2005Packet}.
In contrast, the proposed codec only uses inter-frame prediction for
$Q_{1}$, so the decoder resynchronizes very quickly after packet
loss. Changing the proposed codec to have completely independent packets
would cost approximately 12 bits per frame.

AAC-LD~\cite{Lutzky2005} is another low-delay audio codec whose
total algorithmic delay can range from 20~ms to around 50~ms, depending
on the sampling rate and bit reservoir size. However, its complexity
is higher than that of the proposed codec.

\section{Results and Discussion}

\label{sec:results}

The source code for the proposed codec is available at \url{http://www.celt-codec.org/}
and corresponds to the ``low-complexity mode'' of the CELT codec,
version 0.5.1\footnote{The quality results were obtained using version 0.5.0, which is identical
except for a slightly lower quality VQ search}. Both floating-point and fixed-point implementations are available.

Of the codecs in Section~\ref{sec:related-work}, only ULD's delay
is comparable to the proposed codec's. We include G.722.1C in our
comparison using the highest bitrate available (48~kbit/s) because
of the algorithmic similarities listed earlier, despite its 40~ms
algorithmic delay at 32~kHz. Also, since ULD uses 128-sample frames,
we include a version of the proposed codec with 128-sample frames.
This version uses a 64-sample look-ahead, compared to the 128-sample
look-ahead of ULD. The conditions are summarized in Table~\ref{tab:Characteristics-of-codecs}.

\subsection{Subjective quality}

We use the MUltiple Stimuli with Hidden Reference and Anchor (MUSHRA)
methodology~\cite{BS1534} with 11 listeners\footnote{During post-screening, we discarded results from 3 additional listeners,
who rated on average more than 3/7 samples as 100. We verified that
this post-screening phase did not affect our conclusions. }. We used short excerpts taken from the following material: female
speech (SQAM), pop (Dave Matthews Band, \#41), male speech (SQAM),
harpsichord (Bach), \emph{a cappella} (Suzanne Vega, Tom's Diner),
castanets (SQAM), rock (Duran Duran, Ordinary World), orchestra (Danse
Macabre), and techno. 

\begin{figure*}
\centering\includegraphics[width=0.75\textwidth]{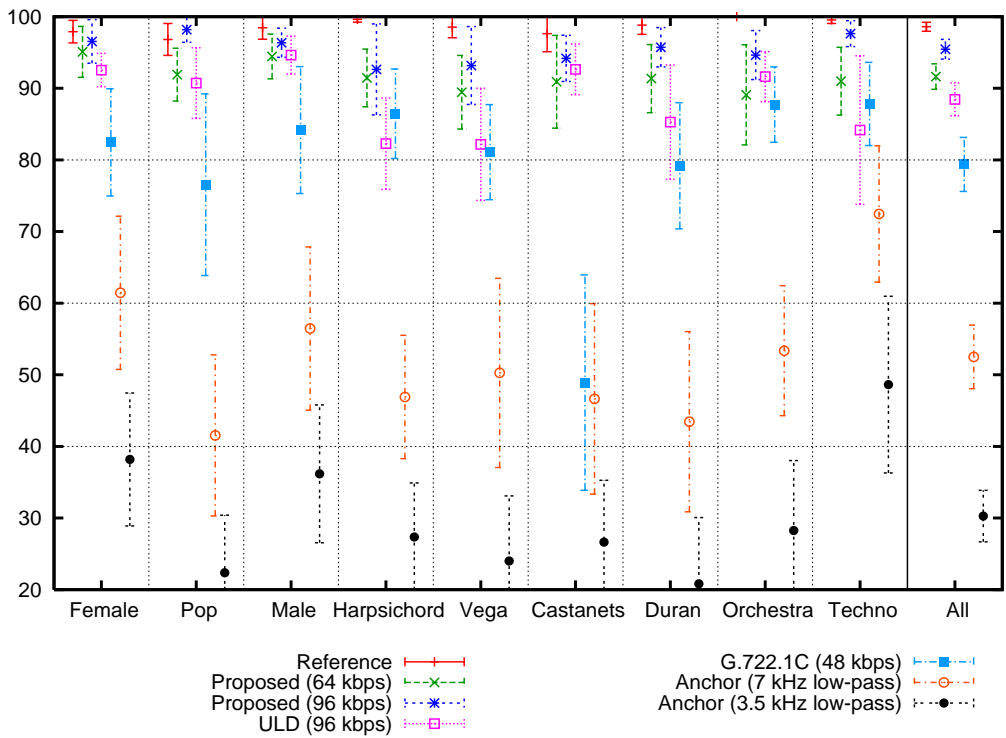}

\caption{Subjective quality of the codecs obtained using the MUSHRA methodology
with 11 listeners. The 95\% non-paired confidence intervals are included.\label{fig:Subjective-quality}}
\end{figure*}

The results are shown in Fig.~\ref{fig:Subjective-quality}. Although
some of the non-paired confidence intervals in Fig.~\ref{fig:Subjective-quality}
overlap, a paired $t$-test reveals higher than 99\% confidence in
all the differences ($P<0.01$). The proposed codec at 64~kbit/s
was better than ULD at 96~kbit/s, although it used a slightly higher
delay. When using a slightly lower delay than ULD and the same 96~kbit/s
bitrate, the proposed codec was clearly better than all other codecs
and configurations tested. %
G.722.1C had the lowest quality, which was expected because its bitrate
is limited to 48~kbit/s.

Unsurprisingly, all ultra low-delay codecs do very well on castanets,
unlike G.722.1C, which has much longer frames. On the other hand,
the highly tonal harpsichord sample was difficult to encode for the
low delay codecs and only the proposed codec achieved quality close
to the reference. G.722.1C did well on the harpsichord due to its
longer frames, despite its lower bitrate.

\begin{figure}
\centering\includegraphics[width=1\columnwidth]{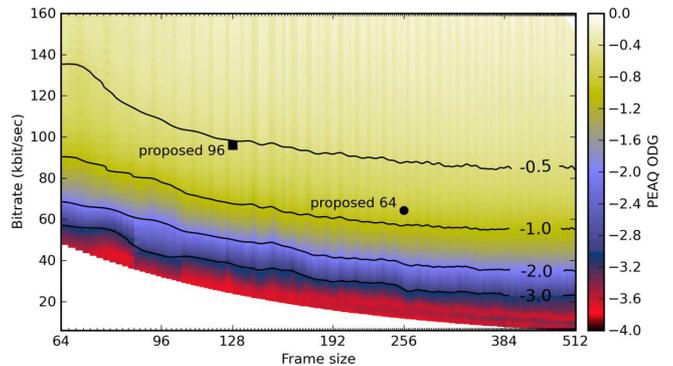}

\caption{Objective Quality Degradation (ODG) measured for different frame sizes.
Equal-quality contours are shown for ODG values -0.5, -1.0, -2.0,
-3.0.\label{fig:PEAQ-map}}
\end{figure}

The proposed codec is able to operate with a wide range of frame sizes.
We evaluated the effect of the frame size and bitrate on the audio
quality. Because of the very large number of possible combinations,
we used PQevalAudio\footnote{\url{http://www-mmsp.ece.mcgill.ca/Documents/Software/Packages/AFsp/PQevalAudio.html}},
an implementation of the PEAQ basic model~\cite{BS1387}. As expected,
Fig.~\ref{fig:PEAQ-map} shows that the bitrate required to obtain
a certain level of quality increases as the frame size decreases.
However, we observe that the bitrate difference between two equal-quality
contours is almost constant with respect to the frame size. For example,
reducing the frame size from 256 to 64 samples (from 8~ms total delay
to 2~ms) results in an increase of 30~kbit/s for the same quality. 

\subsection{Complexity}

The total complexity of the algorithm when implemented in fixed-point
is 11~WMOPS for the encoder and 6~WMOPS for the decoder, for a total
of 17~WMOPS\footnote{Measured by running the fixed-point implementation with operators
similar to the ETSI/ITU basicops and with the same weighting.}. The encoder and the decoder states are very small, requiring around
0.5~kByte for both states combined. The total amount of scratch space
required is 7~kBytes. 

When running on a 3~GHz x86 CPU (C~code without any architecture-specific
optimization), the floating-point implementation requires 0.9\% of
one CPU core for real-time encoding and decoding. The memory requirements
for the floating point version are about twice the fixed-point memory
requirements, which still easily fits within the L1 cache of a modern
desktop CPU.

\section{Conclusion}

\label{sec:conclusion}

We have proposed a low-delay audio codec based on the MDCT with very
short frames, using shape-gain quantization to preserve the energy
in critical bands. We have demonstrated that the subjective quality
of the proposed codec is higher than ULD when operating at the same
bitrate (96~kbit/s) and frame size. In addition, with a slightly
higher delay, the proposed codec operating at 64~kbit/s still out-performs
the ULD codec. 

\bibliographystyle{IEEEbib}
\bibliography{celt}

\end{document}